# Comment: Struggles with Survey Weighting and Regression Modeling

**Roderick J. Little**

I appreciate the opportunity to comment on Andrew Gelman's interesting paper. As an admirer of Gelman's work, it is a pleasure to read his take on the topic of survey weighting, which I have always found fascinating. Since I support Gelman's general approach, I focus on reinforcing some points in the article and commenting on some of the modeling issues he raises.

As a student of statistics, I first encountered weights as the inverse of the residual variance for handling nonconstant variance in regression. I then had a course on sample surveys, where the weights were the inverse of the probability of selection. When these two sets of weights are different, which should be used? This question remained a mystery for many years, and only later did I come to appreciate that it reflects fundamental philosophical differences of design-based versus model-based survey inference.

The design-based approach treats the survey outcomes as fixed, with randomness arising from the distribution of sample selection. Sampling weights, defined as the inverse of the probability of selection, play a pivotal role in design-based inference in yielding estimates that are design unbiased or consistent. Similarly with poststratification, the weight is proportional to the ratio of population and sample counts in the poststrata, and as such involves the distribution of the sample counts rather than outcomes. If the "probability of selection" is replaced by the "probability of inclusion," then nonresponse weighting also enters the picture as the inverse of the estimated probability of response given selection.

The regression approach is model-based, and puts the emphasis on predicting values for nonsampled units in the population. Gelman uses the Bayesian paradigm to generate predictions, but to me the key issue is whether the objective is viewed as prediction. The Bayesian paradigm seems to me (and I think to Gelman) the most natural and compelling framework for prediction (Little, 2004, 2006), but in many situations one can get quite far with likelihood-based methods that do not explicitly add a prior distribution. In summary:

$$\text{design-based} = \text{weighting};$$
$$\text{model-based} = \text{prediction}.$$

This statement is an oversimplification. Design-based weights arise in the context of particular prediction models, so the approaches intersect. A simple example is the stratified mean for stratified samples, which arises as the prediction estimate for a regression on dummy variables for strata. More generally, Little (1991) provides an approximate Bayesian interpretation of design-weighted estimates of regression parameters. Prediction and weighting can be combined, and hybrid approaches are increasingly popular. In particular, Särndal, Swensson and Wretman (1992) take the prediction estimate from a model and then calibrate it by adding weighted sums of residuals, to yield protection against model misspecification. Robins and colleagues (Scharfstein, Rotnitzky and Robins, 1999; Bang and Robins, 2005) use the term "doubly-robust" to describe such estimators, and have popularized them in the general statistics literature; I would be interested in Gelman's views on this alternative approach. My own view is that robustness can be achieved within a pure prediction paradigm by judicious choice of model; see Firth and Bennett (1998), Little (2004) and Little and Zheng (2007).

Design weighting, as represented by the Horvitz–Thompson (HT) estimator and variants, has the virtue of simplicity, and by avoiding an explicit model it has an aura of robustness to model misspecification. It is the "granddaddy of doubly-robust estimators," since it is a prediction estimator for a model

*Roderick J. Little is Professor, Department of Biostatistics, University of Michigan, Ann Arbor, Michigan 48109-2029, USA e-mail: rlittle@umich.edu.*

where the ratios of outcomes to selection probabilities are exchangeable, and it is consistent when either this model or the weights are correctly specified (Firth and Bennett, 1998). However, unthinking application of the HT estimator is dangerous, since inferences based on it can be poor if the underlying HT model is not reasonable. An extreme parody is Basu's (1971) famous elephant example. In work with Hui Zheng, I compared the HT estimator with prediction based on a robust regression model where the relationship between the outcome and the selection probabilities is modeled via a penalized spline. The prediction estimators perform similarly to HT when the HT model is true, and much better when the HT model is violated, in terms of both efficiency and confidence coverage (Zheng and Little, 2003, 2004, 2005). Similar gains in the case of nonresponse are reported in Yuan and Little (2007a, b).

The limitations of design weighting are well illustrated in the case of poststratification considered by Gelman. The design-weighted estimator of the population mean of a variable $Y$ is

$$(1) \quad \bar{y}_w = \sum_{j=1}^{J} P_j \bar{y}_j = \sum_{j=1}^{J} w_j n_j \bar{y}_j \Big/ \sum_{j=1}^{J} w_j n_j,$$

where $P_j, n_j, \bar{y}_j$ are respectively the population proportion, sample count and sample mean in poststratum $j$, and $w_j$ is the design weight, $j = 1, \ldots, J$. From the prediction perspective, $P_j$ is known, and $\bar{y}_j$ is an estimate of the population mean $\bar{Y}_j$ in poststratum $j$. Equation (1) is a model-based estimator for a model that assumes distinct and a priori independent means for each $j$. This estimator works well in large samples, but breaks down if the sample sizes in certain poststrata are small—clearly it totally fails if there are cells where the population proportion is nonzero and the sample size is zero. The prediction approach replaces the estimate $\bar{y}_j$ from the saturated model by an estimate from a more parsimonious model, that is,

$$(2) \quad \bar{y}_{\text{mod}} = \sum_{j=1}^{J} P_j \hat{\mu}_j,$$

where $\hat{\mu}_j$ is the model prediction for poststratum $j$. Note that the "weight" given to the predicted mean $\hat{\mu}_j$ in poststratum $j$ remains the population proportion $P_j$, which seems entirely appropriate because this quantity is *known*. It is the prediction of the poststratum mean $\bar{y}_j$ that is modified, since that is where borrowing strength from data in other cells is needed. This is not possible under a strict design-based approach, because it requires a model for the outcome $Y$. I think that tinkering with the design weights—for example by collapsing poststrata so that they have cases in the sample, or not letting the design weights get too large—puts the emphasis in the wrong place, the weight assigned to the observations, rather than the right place, the predictions of $Y$ in the poststrata. In particular, collapsing over a set of cells assumes an implicit model that the mean of $Y$ is constant in those cells. A more empirical approach to collapsing would be to base the collapsed poststrata on a regression tree model. See Little (1993) for other collapsing ideas.

Gelman replaces the predictions (1) from the saturated model with predictions (2) from a hierarchical regression model. He proposes models that treat main effects in the model as fixed effects by assigning them flat prior distributions, and shrink the interactions toward zero by modeling them with proper prior distributions. This approach provides a good example of the power and flexibility of the Bayesian approach—more a principled extension of design weighting, rather than an alternative. Concerning Gelman's modeling questions, I have the following comments:

1. Gelman writes that "Regression modeling is a potentially attractive alternative to weighting. In practice, however, the potential for large numbers of interactions can make regression adjustments highly variable." However, note that when the strata are based on the joint distribution of design variables, the weighting estimate (1) is based on the saturated model that includes all interactions between these variables, so the weighting approach has this problem in its most extreme form. Any regression model that removes or smooths over interactions should have better precision.

2. If all main effects and interactions appear in the hierarchical model, the resulting estimate of the mean is design consistent, in that it converges to (1) as the sample size increases. The Bayesian approach provides a principled approach to smoothing in small samples.

3. For estimates of means, it is important to model carefully the relationship between the outcome and the propensity to be included (Little, 1983; Rubin, 1983, 1985; Rizzo 1992), and less important to get the relationship with other variables right, since conditional on the propensity, the distributions of other variables are balanced for included and excluded cases by the balancing property of the propensity score (Rosenbaum and Rubin, 1983). This idea motivates penalized spline of propensity prediction (PSPP, Little and An, 2004), which models the relationship between the outcome and the propensity to respond by a penalized spline, and then adds other variables parametrically. Zhang and Little (2005) discuss a simplification of the method, and extensions to parameters other than unconditional means. These methods are formulated for the case of nonresponse propensities, but are readily applied in the setting of other forms of selection, including sample selection.

4. The regression approach conditions on the variables that enter into the weight; hence effects of other variables in the regression model are adjusted for the design variables. To obtain valid unadjusted effects the design variables have to be averaged out. In Section 1.4, Gelman describes this averaging process for models involving interactions, but he does not appear to average over the design variables $X$ when fitting the additive model in Table 1. If I understand his description correctly, then the estimates of change being compared in Table 1 are not comparable, since the regression estimate is adjusted for $X$ and the weighted estimate of change is not. This might account for the differences.

5. Gelman bases variance estimates on the posterior distribution from his Bayesian analysis. These estimates are potentially sensitive to misspecified variance assumptions in the regression models—for example, many survey variables are positive and have a variance that tends to increase with the mean. Assuming standard models with nonconstant variance can lead to incorrect confidence coverage (see, e.g., Yuan and Little, 2007a, b). So, good confidence coverage requires attention to the variance structure as well as the mean structure. One way of avoiding these problems (at the expense of inferential impurity) is to compute sample reuse variance estimates like the bootstrap, and I would be interested in Gelman's attitude to such approaches.

Gelman states that "it is not generally clear how to apply weights to more complicated estimands such as regression coefficients." A wide class of weighted estimates can be obtained from estimating equations where the sample units are weighted by the inverse of their inclusion probabilities (Binder, 1983; Godambe and Thompson, 1986). A special case is weighted pseudo-likelihood, where the estimating equations are derivatives of the log likelihood. However, generating a weighted approximation to the estimating equations for the whole population does not address the problems with weighting discussed above in the poststratification setting.